\documentclass[aps,prf,amssymb,onecolumn,notitlepage,author-year,showkeys,a4paper,preprint]{revtex4-1}
\pdfoutput=1 
\usepackage{color,graphicx}
\usepackage{geometry}
\usepackage{times}
\usepackage{fdsymbol}
\usepackage{afterpage}
\usepackage{pstricks}
\usepackage{tikz}
\usepackage[]{hyperref}
\usepackage{picture}
\hypersetup{colorlinks=true, linkcolor=blue, citecolor=blue, bookmarksopen=true, pdfstartview=FitH}

\usepackage{float}
\usepackage{upgreek}
\usepackage{caption}
\usepackage{amsbsy}
\usepackage{wasysym}
\usepackage{marvosym}
\usepackage{natbib}
\usepackage{amsmath}
\usepackage{lineno}
\begin{document}

\title[Structural inclination angle of scalar]{Structural inclination angle of scalar fluctuations in a turbulent boundary layer}

\author{Krishna M. Talluru} \email{murali.talluru@sydney.edu.au}
\author{Kapil A. Chauhan} \email{kapil.chauhan@sydney.edu.au}

\affiliation{$^\ast$ School of Aerospace, Mechanical and Mechatronic Engineering, The University of Sydney, NSW 2006, Australia\\
	$^\dagger$ School of Civil Engineering, The University of Sydney, NSW 2006, Australia}

\date{\today}
\begin{abstract}
Two-point measurements of concentration are obtained in a meandering passive-scalar plume released at five different heights within the fully-turbulent region of a high-Reynolds number turbulent boundary layer (TBL). These measurements are obtained using a combination of two photo-ionisation detectors (PIDs). A stationary PID is positioned at the plume centreline, while the second PID traverses across the plume in the wall-normal direction. Similar to the large-scale coherence observed in the two-point correlation map of streamwise velocity fluctuations \citep[e.g. Marusic et al.][]{marusic2007reynolds}, the two-point correlation results of concentration also indicate coherence in the scalar field. Particularly, the iso-contours of correlation indicate an inclination of scalar structures in the direction of the flow. The corresponding inclination angle is found to be a weak function of the location of the plume within the boundary layer. However, the inclination angle is observed to be invariant near 30$^\circ$ for all source heights that are below $z/\delta \leq 0.33$. This observed steepness of inclination angle relative to the inclination angle of streamwise velocity is consistently explained through the physical model put forth by Talluru et al. \cite{Talluru2018} that explains the organisation of scalar around the low- and high-speed regions in the flow which inherently also have a characteristic inclination. 
\end{abstract}
\keywords{Two-point correlation, Meandering plume, Structure inclination angle}
\maketitle
\section{Introduction} 	
The local concentration of a scalar (pollutant, heat, etc.) in a turbulent flow exhibits complex features over a broad range of length and time scales. As such, turbulent advection is crucial in most natural and engineering processes such as atmospheric dispersion, mixing and combustion. Although the underlying interaction between the turbulent velocity field and scalar fluctuations is not linear, there are many similarities between the statistics of scalar fluctuations and those of velocity, e.g. in a heated turbulent jet \citep{Antonia1983}. The possibility of exploiting any similarity between velocity and scalar fields in wall-bounded turbulent flows is also of practical significance, for example, prediction of ground level concentration of a pollutant emitted from an elevated source and vice-versa. Thus, many studies are aimed at characterising the role of coherent features of velocity in the dispersion and transport of a scalar \citep[e.g.][]{Talluru2018, Vanderwel2016}. When contaminants are released into a turbulent boundary layer (TBL), the pollutants are mixed by the turbulent eddies in the flow. This mixing leads to strong fluctuations in the concentration of pollutant which are of the same order of magnitude as the mean concentration \citep{Csanady1973}. Such extreme fluctuations are of significant concern when the pollutant is a highly toxic or flammable material. Most dispersion models are capable of predicting the mean pollutant concentration, but provide no or inaccurate information about the amplitude of fluctuations about the mean. Moreover, as the existing models are developed based on single-point statistics, the models cannot account for the spatial coherence of scalar, and to predict concentration at any point in space based on the information of concentration at a reference or measurement location. The present study aims to fill this gap by characterising the spatio-temporal correlation of concentration fluctuations in the vertical direction using experimental data. Specifically, this study focuses on the near-field behaviour of a passive plume, i.e., the region, where the plume meanders. First, a brief discussion is presented here on the two well-accepted results in turbulent boundary layers and plume dispersion that form the basis of the current study. Then we briefly outline our experiments of two-point measurements of instantaneous concentration in a point-source scalar plume. Results of statistics, two-point correlation and observed coherence follow thereafter.
\subsection{Inclination angle of streamwise velocity fluctuations in a turbulent boundary layer}
Inclined structures, typically attached to the wall, are well-known features of streamwise velocity fluctuations in a turbulent boundary layer. One of the earliest studies that reported this behaviour is the flow visualisation study in a zero pressure gradient TBL \cite{head1981new}. They reported that at moderately high Reynolds number ($Re_\theta > 2000$), the TBL consisted of elongated hairpin vortices, inclined at a characteristic angle of approximately 45$^\circ$ to the wall. They suggested that the hairpin vortices could organise themselves at high Reynolds numbers to form larger structures that are typically oriented at an angle of 20$^\circ$ to the wall. The view of hairpin packets was further confirmed  in the high-resolution PIV study of a turbulent boundary layer across a range of Reynolds number (930 $ \leq Re_\theta \leq$ 6845) \cite{Adrian2000}. They observed the inclined coherent features in the streamwise-wall-normal plane across the entire range of $Re_\theta$. Other studies \cite{ganapathisubramani2003characteristics, ganapathisubramani2005investigation, hutchins2005inclined} in laboratory turbulent boundary layers also indicated that the average coherent structure in the logarithmic region has a low inclination angle made up of organised packets of attached eddies. A more conclusive evidence was provided by \citep[e.g. Marusic et al.][]{marusic2007reynolds}, who showed that the two-point correlation map of streamwise velocity fluctuations obtained in laboratory and atmospheric boundary layers presented a characteristic inclination angle of $14^\circ$. Very importantly, they observed that the inclination angle remained invariant over three orders of magnitude change in Reynolds number, $10^3 \leq Re_\tau \leq 10^6$.   
\subsection{Meandering of a slender passive scalar plume}
Observations of a smokestack plume in the atmosphere show that its instantaneous structure is puffy with narrow sections scattered between wide ones, and the plume meanders in a highly variable manner \citep{Fernando2012}. This variability is caused by the turbulent wind field in which the plume is fully embedded, and the scale of atmospheric turbulence that influences the plume fluctuations will change as the plume size increases \cite{sawford1985lagrangian}. For example, in the near-field region, when the plume is smaller than the turbulent eddies, the plume is swept up and down about the plume axis by the turbulent eddies, but when the plume is larger than the eddies, i.e. in the far-field region, the eddies no longer advect the  plume but merely cause minor fluctuations deep within the plume \cite{hanna1989time, Yee1993}. The meandering of plume results in the intermittent high and low concentrations in the time series signal obtained at a single point in space as observed in laboratory wind tunnel \citep{Fackrell1982, Nironi2015} and full-scale atmospheric \citep{Yee1993} measurements. Recently, \citep[Talluru et al.][]{Talluru2018} conducted a study of point source plumes released at eight different elevations in a TBL and presented a physical model to explain the relative organisation of the large-scale coherent velocity structures and the concentration fluctuations in the plume. Through flow visualisation as well as joint-pdfs (probability density functions) and cross-spectra analysis, they comprehensively showed that the large-scale turbulent eddies cause meandering in the near-field region of the plume. Talluru et al. \citep{Talluru2018} also observed that above the plume centreline, low-speed structures have a lead over the meandering plume, while high-momentum regions are seen to lag behind the plume below its centreline. 

\par A natural question that arises here is whether the structure of scalar fluctuations is similar to that of inclined velocity structures in a TBL, and secondly, how that structure varies as the plume size increases from the near-field to the far-field region. The latter question is partly addressed in a recent numerical study of scalar dispersion in a turbulent channel flow \cite{srinivasan2011direction}. They observed that the concentration fluctuations have a similar behaviour as that of velocity structures in the far-field region. However, little is known about the organisation of scalar structures in the near-field region. To this end, two-point measurements of concentration in the near-field region of a passive scalar plume are analysed in the present study.    
\section{Experimental details}
\begin{figure}
	\centering
	\includegraphics[scale=0.625]{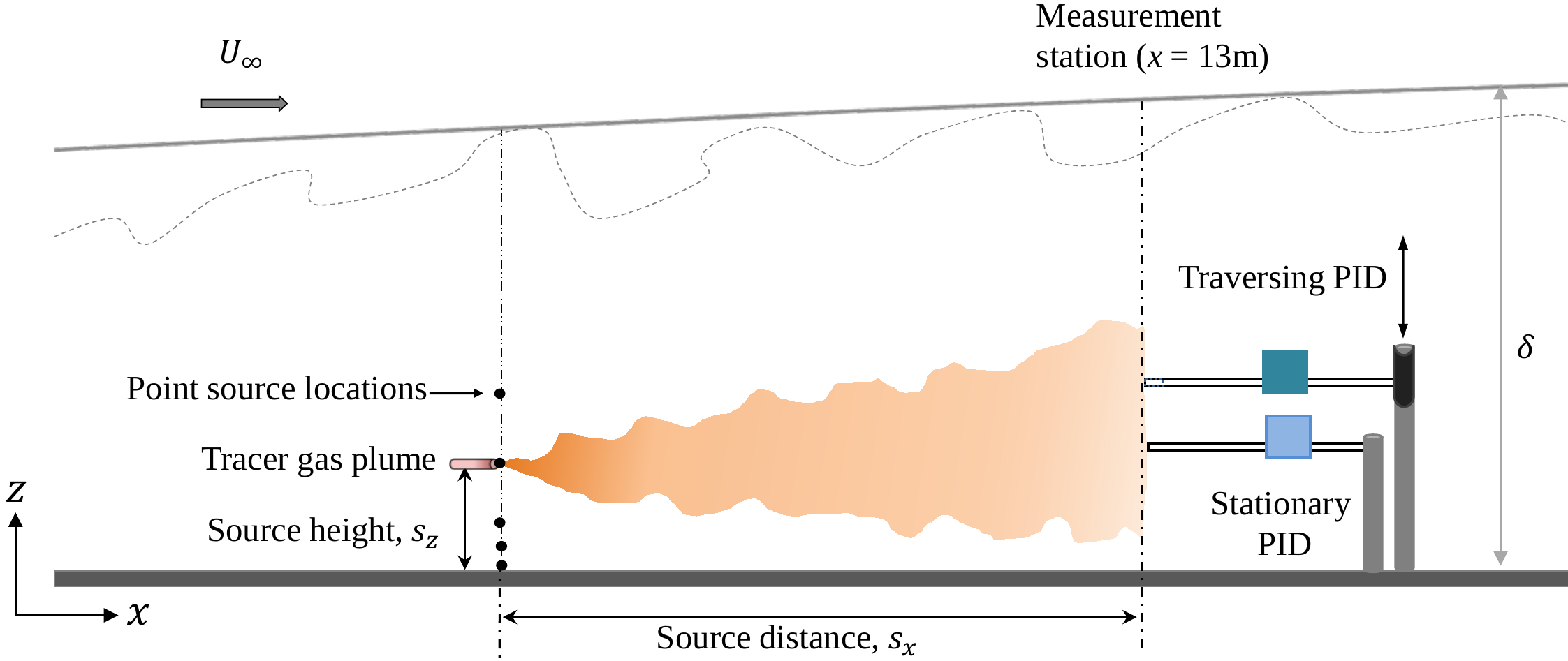}\\
	\caption{Schematic of the experimental set-up used for obtaining two-point measurements of concentration in the vertical direction across the plume.}
	\label{fig:schematic}
\end{figure}
Experiments are performed in the boundary layer wind tunnel at the University of Sydney. Full details of the facility are given in \cite{Talluru2017a} and a schematic of the test rig is shown in Fig.~\ref{fig:schematic}. In the present study, measurements are conducted in a turbulent boundary layer developing over a streamwise fetch of 13\,m at a nominal free stream velocity ($U_\infty$) of 10\,m/s. The boundary-layer thickness ($\delta$) and the friction velocity ($U_\tau$) at the measurement location are approximately 0.31\,m and 0.37\,m/s, respectively, yielding a friction Reynolds number, $Re_\tau \equiv \delta U_\tau/\nu \approx$ 7850 ($\nu$ is the kinematic viscosity of air). Here, upper-case (e.g., $C$) and lower-case letters (e.g., $c$) indicate mean and fluctuating quantities, respectively and $z$ represents the wall-normal coordinate. A point source tracer gas (a mixture of 1.5\% iso-butylene and 98.5\% Nitrogen) is released at five different heights, $s_z/\delta$ = 0.004, 0.044, 0.1, 0.25 and 0.33, at a streamwise distance $s_x/\delta = 1$ upstream of the measurement location, specifically to investigate the near-field behaviour of the plume. The tracer gas is released iso-kinetically (i.e. the source velocity is matched with the local velocity of flow) into the turbulent boundary layer. As shown in the schematic, a stationary PID is positioned at the source-height, $z = s_z$, while a second PID is traversed along $z$. The spanwise gap between the two PIDs is approximately 3\,mm, in accordance with the results of \cite{Metzger2003}, who showed that a PID has negligible influence on an adjacent probe when separated by a distance greater than 3$d$, where $d$ = 0.76\,mm is the inlet diameter of PID probes used in this study. Details of calibration procedure, spatial resolution and frequency response of PID are also given in \cite{Talluru2017b, Talluru2019a}. Although the experimental test-rig and the measurement procedures used in this study are similar to that of \cite{Talluru2017a}, the present study is distinct and reports simultaneous measurements of concentration at two points in the plume. To our knowledge, such measurements using point-detection probes have not been reported before.
\section{Mean and r.m.s profiles of concentration}
\begin{figure}
	\centering
	\includegraphics[scale=1]{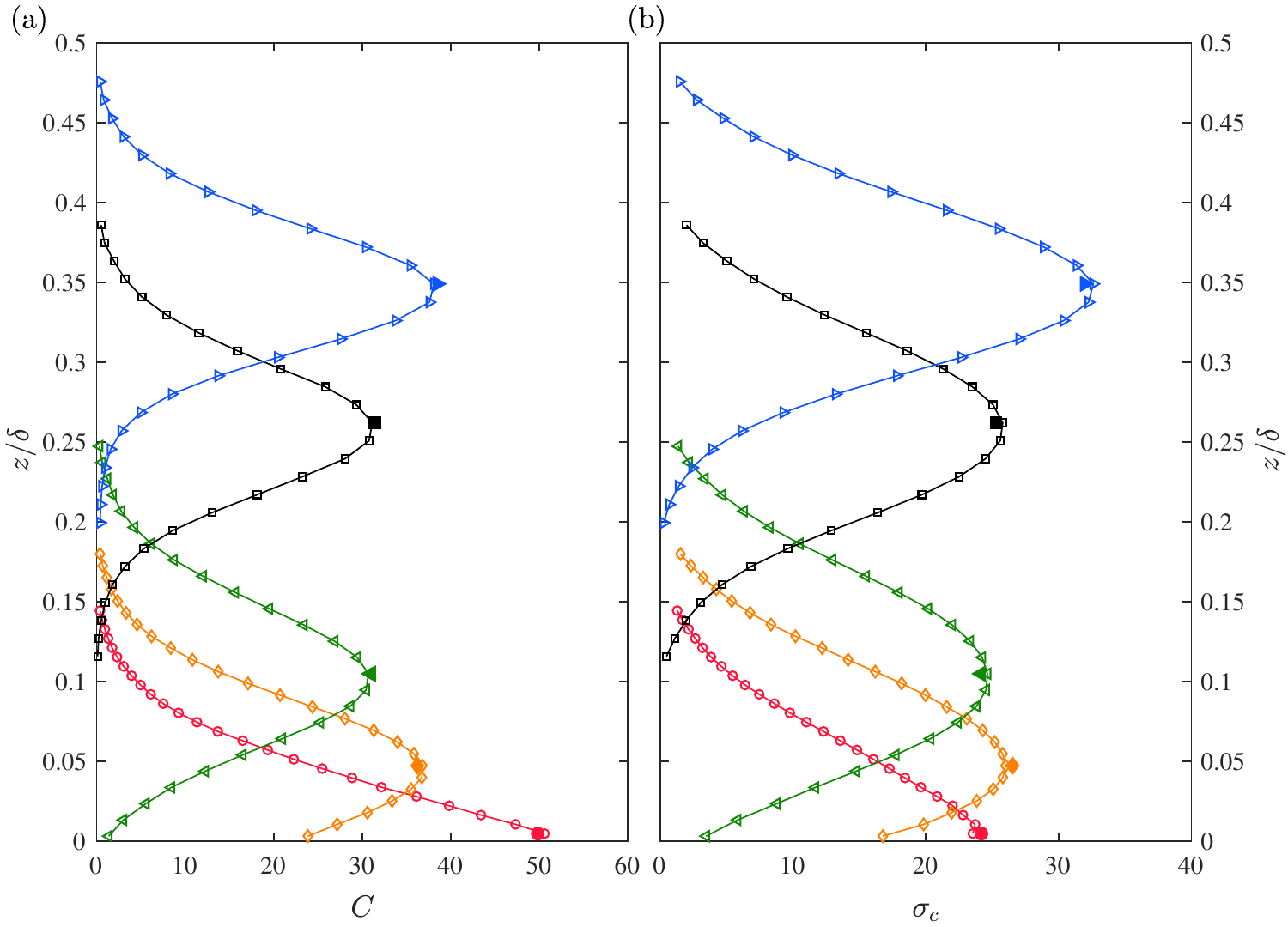}\\
	\caption{Profiles of (a) mean $C(z)$, and (b) r.m.s $\sigma_c(z)$ of concentration in the vertical direction. Open symbols represent the data from the traversing PID in a point source plume released at different source heights; $s_z/\delta$ = 0.004  (\textcolor{red}{$\ocircle$}); 0.044 (\textcolor{orange}{$\lozenge$}); 0.1 (\textcolor{green}{$\lhd$}) ; 0.25 (\textcolor{black}{$\Box$}); 0.33 (\textcolor{blue}{$\rhd$}). Solid symbols represent the measurement from the stationary PID.}
	\label{fig:CmeanCrmsZ}
\end{figure}
\par Figure \ref{fig:CmeanCrmsZ}(a,\,b) plots the mean ($C$) and r.m.s $\sigma_c(z)$ of concentration along $z$ when the plume is released at five heights within the TBL. The open symbols in the figures represent the statistics obtained from the traversing PID, while the solid symbols are from the stationary PID. It is observed that the statistics from both the PID probes agreed very well to within $\pm 2\%$. For a ground level source, $C(z)$ is found to have an exponential form (Eqn.~\ref{eqn:exp}), whereas the elevated sources exhibit a reflected Gaussian behaviour (Eqn.~\ref{eqn:RG}), originally proposed by Fackrell and Robins \cite{Fackrell1982}. These results are consistent with those of Fackrell and Robins \cite{Fackrell1982} and Nironi et al. \cite{Nironi2015}. On the other hand, the root mean square (r.m.s) of concentration fluctuations ($\sigma_c$) along $z$ has a simple Gaussian behaviour (Eqn.~\ref{eqn:SG}) in $z$ for all sources. Note that $\delta_{z}$ and $\sigma_z$ are the plume half-widths based on mean concentration and variance, respectively. In general, half plume-width is defined as the distance from plume centreline, where the maximum drops to half its value. A closer look at the mean concentration profiles reveals that $\delta_z$ and $\sigma_{z}$ vary with source height. Overall, it is clearly evident from Figs.~\ref{fig:CmeanCrmsZ}(a, b) that the behaviour of mean and r.m.s. of concentration is inherently different from the statistical behaviour of the streamwise velocity. 
\begin{align}
\frac{C(z)}{C_{\mathrm{max}}} & \sim  \exp\bigg\{\Big(\frac{z}{{\delta_z}}\Big)^{\alpha}\bigg\}
\label{eqn:exp}\\
\frac{C(z)}{C_{\mathrm{max}}}  & \sim \exp\bigg\{-\mathrm{ln2}\Big(\frac{z - s_z}{\delta_{z}}\Big)^{2}\bigg\} +\exp\bigg\{-\mathrm{ln2}\Big(\frac{z + s_z}{\delta_{z}}\Big)^2\bigg\} \label{eqn:RG}\\
\frac{\sigma_c(z)}{{\sigma_c}_{\mathrm{max}}}  & \sim \exp\bigg\{-\mathrm{ln2}\Big(\frac{z - s_z}{\sigma_{z}}\Big)^{2}\bigg\}
\label{eqn:SG}
\end{align}
\par The physical mechanism for the Gaussian behaviour of a point-source scalar plume has been explained through the meandering nature of the plume \citep{Gifford1959}. Although no similarity exists between the mean of concentration and velocity (or moments of their fluctuations), it is well-understood that a passive scalar plume meanders due to separation between the length-scales of the turbulent flow and the characteristic length-scale of the plume, i.e. dissimilarity between the integral length scale of velocity fluctuations and the plume width or the source diameter \citep{Fackrell1982, Karnik1989, hanna1989time}. Since the integral length scales of a turbulent flow characterise the extent of coherent motions, it is anticipated that the coherent structures in the velocity will impose the instantaneous organisation of concentration field. In the remainder of this paper, we will identify and then characterise such organisation of scalar fluctuations in the streamwise/wall-normal plane, particularly for the inclination in the mean flow direction.
\section{Single- and two-point correlation functions}
\begin{figure}
	\centering
	\includegraphics[scale=1]{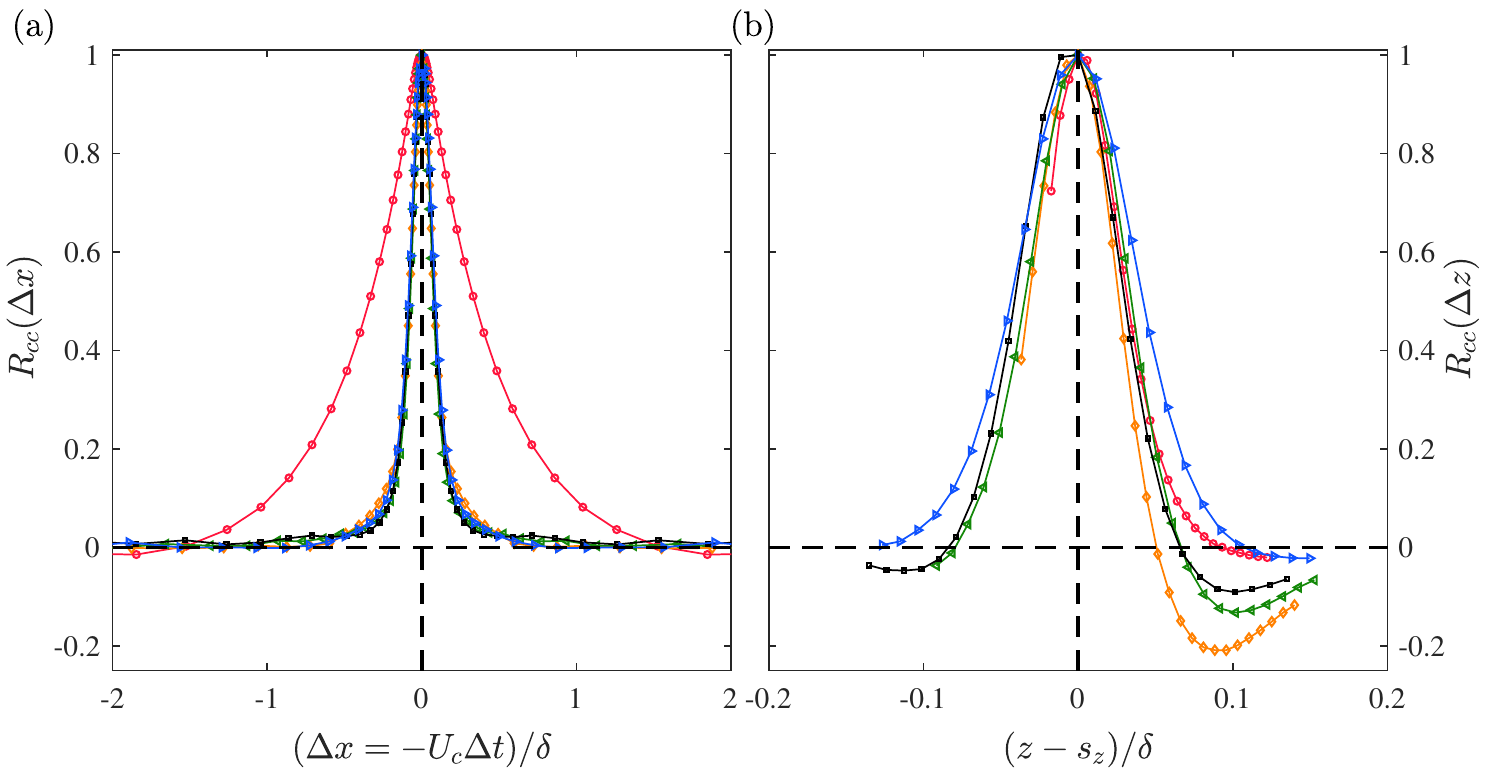}
	\caption{Autocorrelation functions in the (a) streamwise and the (b) wall-normal directions computed using Eqns.~(\ref{eq:Rcc1}) and (\ref{eq:Rcc2}), respectively. Symbols same as in Fig.~\ref{fig:CmeanCrmsZ}.}
	\label{fig:Correlation}
\end{figure}

Using the time-series information of concentration fluctuations at two distinct points, the autocorrelation functions along the streamwise and the wall-normal directions are computed using Eqns.~(\ref{eq:Rcc1}) and (\ref{eq:Rcc2}), respectively. The Eqn.~(\ref{eq:Rcc2}) represents the correlation at zero time-shift between the signals at the plume centreline (stationary PID) and at another $z$-location (traversing PID). The results of spatial autocorrelation functions (computed with reference to the plume source height) are plotted in Figs.~\ref{fig:Correlation}(a,\,b). The temporal domain is converted to the spatial domain using the relationship $\Delta x = -U_c \Delta t$, where $\Delta t$ is the time-lag between two concentration signals and $U_c$ is the convection velocity taken to be the mean velocity at the corresponding wall-normal height \footnote{Note that since the plume evolves rapidly in the streamwise direction, particularly close to the source, Taylor's hypothesis of frozen turbulence cannot be applied to the scalar field precisely. However in this study, we aim to compare the inclination angle of scalar coherence with the inclination angle of streamwise velocity fluctuations, where the Taylor's hypothesis is extensively used to convert temporal domain to a spatial one. Thus to make an equivalent comparison the relation, $\Delta x = -U_c \Delta t$ is also used for the scalar field.}. Looking at the results presented in Fig.~\ref{fig:Correlation}(a), it is evident that the correlation sustains for longer distance in the case of a ground level source. The integral length scale (obtained from the autocorrelation function, $R_{cc}$) for the ground level plume is found to be five times larger than that for an elevated plume. This is primarily due to the wall-effect; the wall has a blocking effect on the plume spread resulting in higher concentration near the surface that persists for large distances from the source leading to a gradual drop in the correlation value. On the other hand, the turbulence in the flow causes an elevated plume to sway above and below the centreline causing intermittency which in turn causes the correlation magnitude to drop quickly. Interestingly, the curves of $R_{cc}(s_z,\Delta x)$ are almost identical for all the elevated sources. This observation suggests that at elevated sources the local velocity is the convection velocity for scalar fluctuations. 
\begin{align}
R_{cc}(s_z,\Delta t) & = \frac{\overline{c(s_z,x)\;c(s_z, x+\Delta x)}}{\big[\sigma_c(s_z)\;\sigma_c(s_z)\big]}, \label{eq:Rcc1} \\
R_{cc}(s_z,z) & = \frac{\overline{c(s_z)\;c(z)}}{\big[\sigma_c(s_z)\;\sigma_c(z)\big]}, \label{eq:Rcc2}
\end{align}
In contrast, the behaviour of autocorrelation functions $R_{cc}(s_z,z)$ in the wall-normal direction is not similar for different source heights, as evident in Fig.~\ref{fig:Correlation}(b), which is anticipated since the spread of plume in the wall-normal direction (e.g. the half-width) varies with source height. Note that the wall-normal shift is presented with respect to the source height as $(z-s_z)/\delta$. It is seen that $R_{cc}(s_z,z)$ is not symmetric about the plume centreline especially when the plume spread is influenced by the wall as a result of the non-homogeneity of turbulence in the wall-normal direction. Interestingly, the autocorrelation functions for the plumes ($s_z/\delta$ = 0.044, 0.1 and 0.25) exhibit both positive and negative correlations. This implies that there are instances at which the concentration fluctuations (about the mean value) at the plume centreline and at some distance away from the centreline are of opposite signs. In other words, this result presents a convincing evidence for the meandering behaviour of the plume in the vertical direction. Further, the magnitude of negative correlation drops with increasing source height from $s_z/\delta$ = 0.044 to 0.25, which aligns well with the trend of wall-normal turbulence intensity in a TBL. 
\section{Two-point correlation and inclination angle}
\par
\begin{figure}
	\centering
	\includegraphics[scale=1]{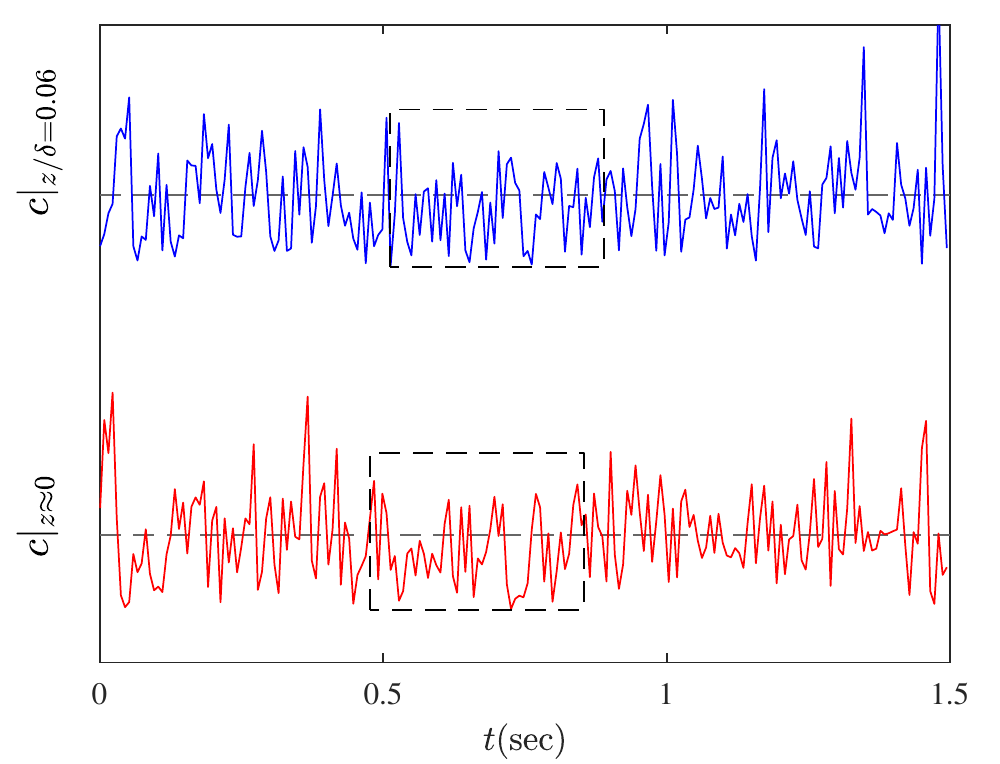}
	\caption{Sample records of simultaneous signals of concentration at the wall and a point in the log-region of turbulent boundary layer. Horizontal dashed lines indicate mean concentration values. The dashed rectangles highlight the similarity in the two signals with a finite time shift.}
	\label{fig:timeseries}
\end{figure}

 Before discussing the two-point correlation map of concentration fluctuations, we will first analyse the time-series signals of concentration for a plume released at $s_z/\delta$ = 0.004, i.e. the ground level source in our measurements. Figure \ref{fig:timeseries} shows a sample of 1.5 second record of concentration measurements at the wall ($z = 0$) and at another location in the log-region of TBL (i.e., $z/\delta = 0.06$). One can readily appreciate the qualitative similarity of these two signals, indicating that there is a strong correlation between their variation, however with a shift in time. The similarity is quantified using the correlation function which also provides an estimate of the time-shift or equivalently the spatial-shift ($\Delta x = -U_c \Delta t$). The correlation coefficient, $R_{cc}$ between two signals is defined as,
\begin{equation}
R_{cc}(z,\Delta x) = \frac{\overline{c(s_z,x)\;c(z, x+\Delta x)}}{\big[\sigma_c(s_z)\;\sigma_c(z)\big]}, \label{eq:Rcc}
\end{equation} 
where overbar indicates time average. 
\par
\begin{figure}
	\centering
	\includegraphics[scale=1]{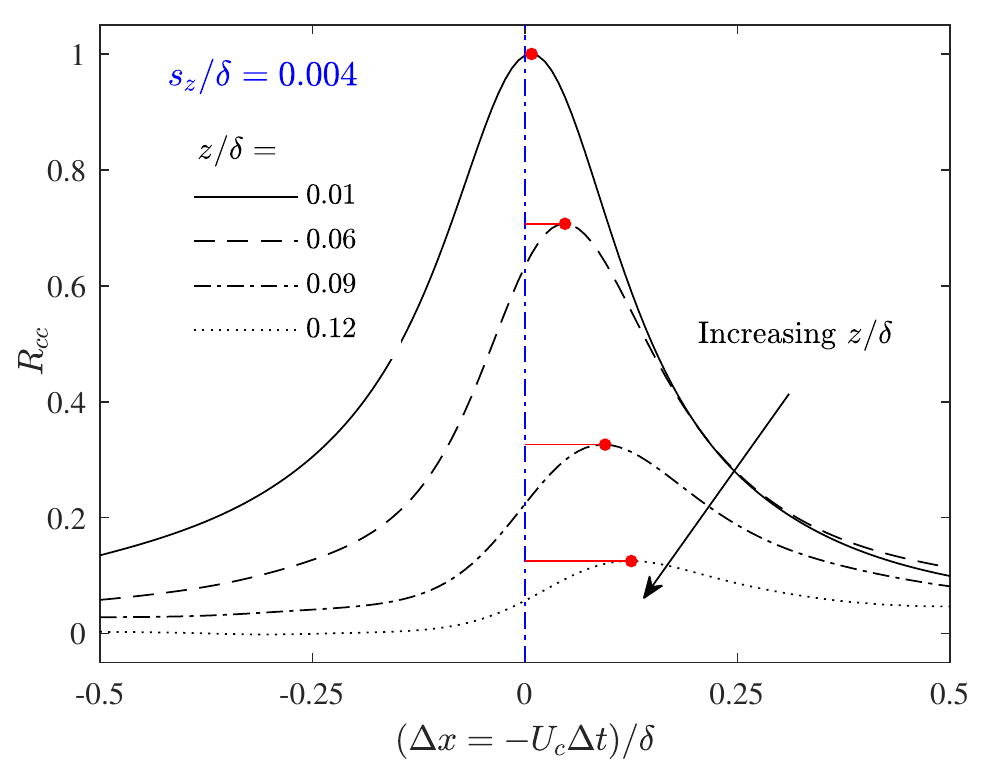}
	\caption{Cross correlation of $c(z=0)$ at the wall and $c(z>0)$ at elevated positions in the plume released at ground-level. Red lines indicate the magnitude of $\Delta x^\ast$.}
	\label{fig:crosscorr}
\end{figure}

\begin{figure}
	\centering
	\includegraphics[scale=1]{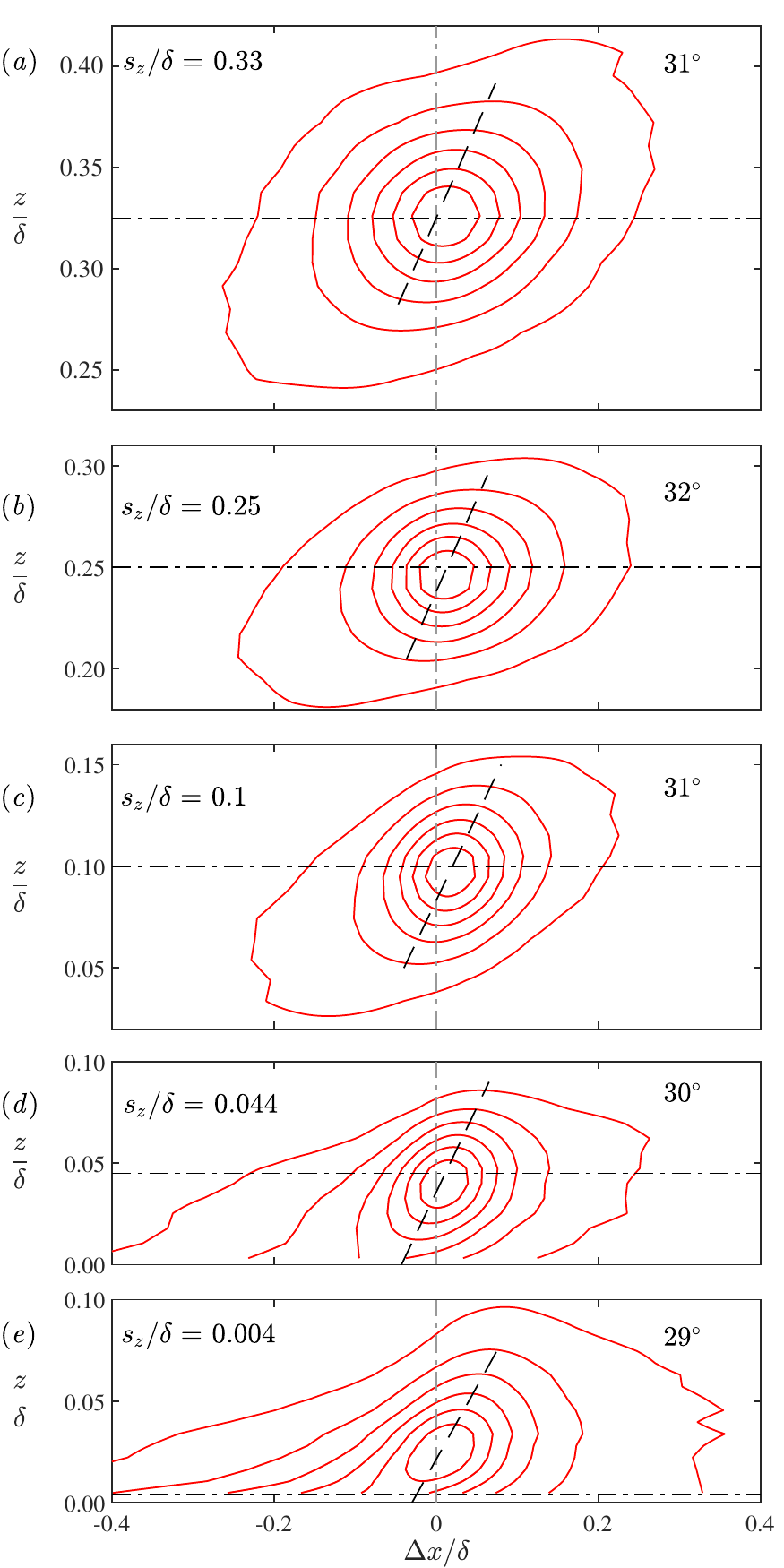}
	\caption{Two dimensional two-point correlation map of concentration fluctuations in the streamwise/wall-normal plane at five source heights, $s_z/\delta = $ (a) 0.33; (b) 0.25; (c) 0.1; (d) 0.045 and (e) 0.004. Contour levels are in increments of 0.15 from 0.1 to 1. The dashed lines represent the inclination angles indicated on the top-right of each sub-plot.}
	\label{fig:Rcc_YZ}
\end{figure}

 The correlation coefficient for the sample signals in Fig.~\ref{fig:timeseries} is plotted in Fig.~\ref{fig:crosscorr}, along with correlation coefficient between $c(z \approx 0)$ and signals at other $z$-locations. High levels of correlation are observed across the plume, however the peak correlation magnitude decreases with increasing distance from the wall. Further, it is seen that with increasing distance from the wall, the location of peak correlation occurs at a non-zero shift on the abscissa. Thus, one can define the inclination angle, $\theta = \mathrm{arctan}[(z-s_z)/\Delta x^\ast]$, where $z$ is the wall-normal position in the plume and $\Delta x^\ast$ is the separation distance corresponding to the peak value in $R_{cc}$. Similarly, the inclination angles inferred from correlation functions at different heights in the plume are calculated, and the average angle for the ground-level source is found to be 29$^\circ$, as indicated in Fig.~\ref{fig:Rcc_YZ}(e) for $s_z/\delta  = 0.004$. The above procedure is repeated for the plume released at four other source locations that span the near wall, log-, and outer-regions of the TBL. Contour maps of correlation coefficient $R_{cc}$ are plotted as a function of $\Delta x/\delta$ and $z/\delta$ in Figs.~\ref{fig:Rcc_YZ}(a-e) for better visualization of the coherence in the scalar field and its inherent inclination.
\par
\begin{figure}
	\centering
	\includegraphics[width=\textwidth]{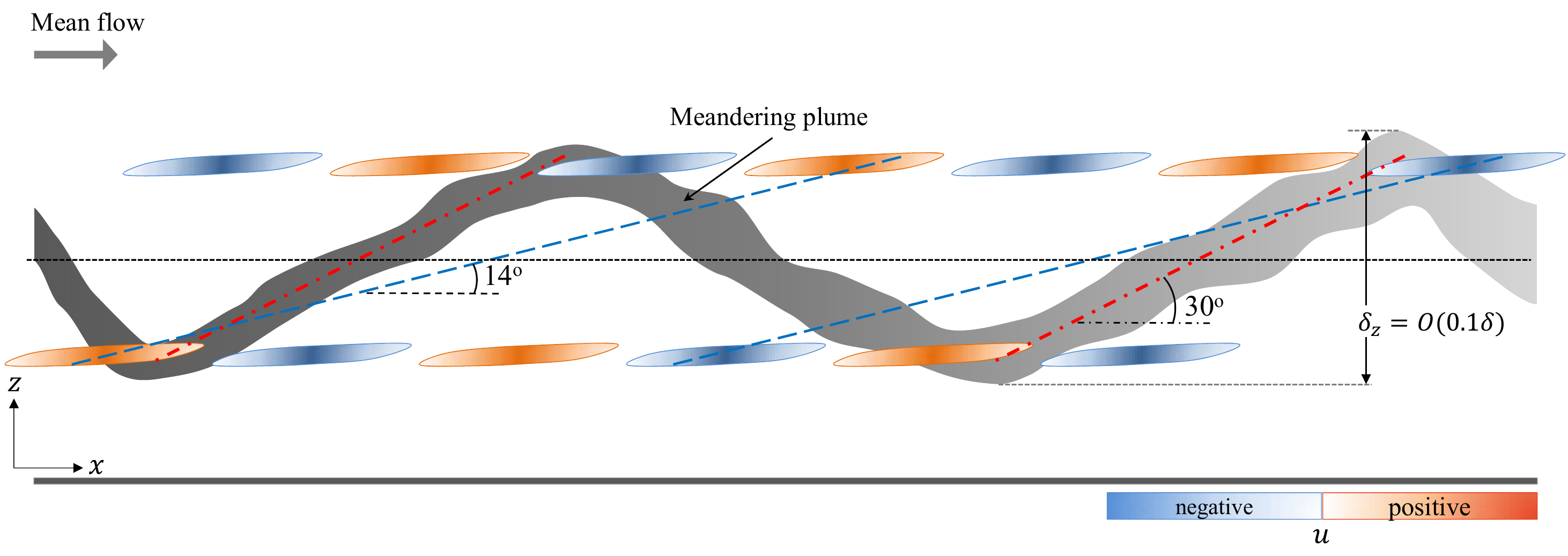}
	\caption{A schematic of the inclination angles of velocity (14$^\circ$, dashed lines) and scalar (30$^\circ$, dot-dashed line) structures in a meandering plume based on the physical model given by \cite{Talluru2018}.}
	\label{fig:PhysicalModel}
\end{figure}

 The nominal inclination angle of scalar structure is found to be approximately $30^\circ$ for plumes released at four other source locations, $s_z/\delta$ = 0.045, 0.1, 0.25 and 0.33.  Note that the height of ordinate axis in each sub-plot is proportional to the plume spread. Some inferences can be made by comparing the average structure of concentration fluctuations against that of velocity structures previously reported in \cite{Hutchins2011,Talluru2014a}. Firstly, these results show that the size of correlated regions scale with $\delta$, suggesting that scalar structure is governed by length scales in the outer-region of a TBL. Recently, \cite{Talluru2018} provided the evidence for the role of large-scale velocity fluctuations in the scalar transport across the TBL through their role in streamwise scalar-flux. Secondly, it is noticed that the positively correlated regions of concentration fluctuations are inclined to the horizontal axis at $\theta\approx$ 30$^\circ$ within the region, $-0.5 \leq (z-s_z)/\delta_z \leq 0.5$, for all sources in the fully turbulent region, i.e. $s_z/\delta \leq 0.33$. This is an unexpected result, as this angle is considerably steeper than the corresponding inclination angle of $\approx14^\circ$ noticed in the correlations of streamwise velocity structures \citep{marusic2007reynolds}. However, the phenomenological model of \cite{Talluru2018} that outlines the preferential organisation of scalar motions with respect to large-scale low- and high-speed structures in the flow explains this observation. A schematic of the physical model is presented in Fig.~\ref{fig:PhysicalModel}, where a meandering plume along with a cluster of low- and high-speed regions are indicated. In this schematic, low-speed or high-speed regions of velocity fluctuations are indicated at $14^\circ$ to the horizontal, as correlation between similarly signed (low- or high-speed) fluctuations will contribute to the 14$^\circ$ angle observed in numerous past studies \citep{Adrian2007, marusic2007reynolds, Hutchins2011, Talluru2014a}. Talluru et al. \cite{Talluru2018} have shown that above the plume centreline, low-speed regions in velocity are correlated with high-concentration scalar regions, resulting in $\overline{uc}<0$, whereas below the centreline, high-speed regions are correlated with the high-concentration region resulting in $\overline{uc}>0$. Thus, instantaneously, even though the velocity fluctuations might have the same sign across the whole plume width (at a nominal angle of 14$^\circ$), the concentration field itself is not consistently correlated with velocity fluctuations. Regions of coherent scalar fluctuations in the meandering plume thereby attains a much steeper inclination angle as shown in figure \ref{fig:PhysicalModel}, that is approximately $30^\circ$, as observed in this study. Thus, we find that although the mean and r.m.s. of scalar fluctuations has a symmetric distribution (Gaussian or Gaussian-like), instantaneously it has a characteristic inclination angle that manifests as the inclination observed for $R_{cc}$. Thus, the degree of meandering is not symmetric about the plume centreline and has a spatial preference.
\par In a recent numerical study of a turbulent channel flow, Srinivasan et al. \cite{srinivasan2011direction} showed that the forward correlations of scalar structures are oriented at a angle of $76^\circ$ with the vertical (equivalent to $14^\circ$ to the horizontal), which is in contrast with the present results. This difference could be attributed to two reasons - (i) how the scalar is released in the flow or the computational domain and (ii) whether the plume is in the near- or far-field region. In our wind tunnel experiments, the scalar is released via a point source and the measurements are performed in the near-field region, where the plume is still meandering. On the other hand, Srinivasan et al. \cite{srinivasan2011direction} studied a flow, where the scalar is released at infinite number of points in the wall-normal plane of the computational domain. This implies that there is no concentration gradient in the vertical direction in the numerical simulations performed by \cite{srinivasan2011direction}, and hence their results are not directly comparable to the present study. We would like to emphasise that the observations made in this study are closely linked to the meandering nature of the plume. It is very likely that experimental measurements further downstream, i.e., in the far-field region, where the plume is thoroughly dispersed, the correlation map of concentration fluctuations will show an inclination angle near $14^\circ$, consistent with the velocity structures and the study of Srinivasan et al. \cite{srinivasan2011direction}.  It would be insightful to further investigate the  inclination angle of scalar structures (released via a point source) at large source distances in the future studies. 
\section{Conclusions}
For the first time, an unique set of two-point concentration measurements in the vertical direction are analysed with the aim of characterising the structure of passive scalar fluctuations in a TBL. For this, experiments are conducted by varying the location of point source at five positions that span the near-wall, log-, and outer-regions of TBL. The separation distance between the source and the measurement location is chosen to be $1\delta$ in order to investigate the near-field meandering behaviour of the plume. Autocorrelation functions computed using single-point measurements at the plume centreline revealed that the correlation is sustained for longer distances in the case of a ground level plume due to the influence of the wall. Further, the concentration fluctuations along $z$ has both positively and negatively correlated regions indicating that the plume meanders in the vertical direction (and likewise in the spanwise plane). Based on the two-point correlation analysis, a relatively steeper structural angle of approximately $30^\circ$ is observed in the scalar structures in the region ($-0.5 \leq (z-s_z)/\delta_z \leq 0.5$) close to the plume axis. These results are found to be entirely consistent with the physical model put forth by \cite{Talluru2018} for the large-scale organisation of scalar concentration relative to the large-scale velocity structures in a TBL, i.e., the preference of scalar to correlate with low- or high-speed regions in a turbulent boundary layer results in the coherent organisation of the scalar structures. 	
\bibliographystyle{spbasic}
\bibliography{Bib_Gaussian}
\end{document}